# NUMERICAL MODEL FOR CONDUCTION-COOLED CURRENT LEAD HEAT LOADS

M. J. White[1], H. D. Brueck[2], and X. L. Wang[2]

[1]Fermi National Accelerator Laboratory (FNAL)
P.O. Box 500, Batavia, IL 60510, U.S.A.

[2]Deutsches Elektronen-Synchrotron (DESY)
Notkestraße 85, 22607, Hamburg, Germany

## ABSTRACT

Current leads are utilized to deliver electrical power from a room temperature junction mounted on the vacuum vessel to a superconducting magnet located within the vacuum space of a cryostat. There are many types of current leads used at laboratories throughout the world; however, conduction-cooled current leads are often chosen for their simplicity and reliability. Conduction-cooled leads have the advantage of using common materials, have no superconducting/normal state transition, and have no boil-off vapor to collect.

This paper presents a numerical model for conduction-cooled current lead heat loads. This model takes into account varying material and fluid thermal properties, varying thicknesses along the length of the lead, heat transfer in the circumferential and longitudinal directions, electrical power dissipation, and the effect of thermal intercepts. The model is validated by comparing the numerical model results to ideal cases where analytical equations are valid. In addition, the XFEL (X-Ray Free Electron Laser) prototype current leads are modeled and compared to the experimental results from testing at DESY's XFEL Magnet Test Stand (XMTS) and Cryomodule Test Bench (CMTB).

## INTRODUCTION

Conduction-cooled current leads are often used at laboratories to power superconducting magnets due to their simplicity and reliability. Conduction-cooled leads utilize common materials with no superconducting/normal state transition and have no boil-off vapor to collect or the associated liquefier load on the refrigeration system. Improperly sized conduction-cooled current leads can lead to significant static and/or dynamic heat loads. Analytical methods of estimating heat loads yield fast and straightforward solutions, but neglect important heat transfer processes within the current leads. A simple analytical model will be presented and compared to a numerical model written to predict conduction-cooled current lead heat loads using idealized conditions. In addition, the numerical model will be compared to experimental results from testing prototype XFEL current leads. The models presented in this paper are intended to be general in nature, but the parameters for the current leads tested at DESY in the prototype cryomodule PXFEL2 will be used to compare the models.

The PXFEL2 leads are adapted from a CERN design [1, 2]. The leads have an 85Cu/15Zn brass core with a diameter of 3 mm and an overall length of 1.280 m. The first 0.500 m of the lead and the last 0.555 m of the lead have an electroplated copper thickness of 0.6 mm, while the region in the middle of the lead has a copper thickness of 0.13 mm. Outside the copper are 5 electrically insulating layers of Kapton® polymide film with a total thickness of 0.135 mm, which prevents any electrical conduction through the stainless steel tube. The core is inserted into a stainless steel tube with an outer diameter of 5.6 mm, a thickness of 0.2 mm and a length of 1.200 m. The radial distance for the helium filled gap between the polymide film and the stainless steel is 0.2 mm for the 0.6 mm thick sections of copper and 0.38 mm for the 0.13 mm thick section of copper. The helium in the gap space is at the same pressure as the helium bath. The first 20 mm of the lead length protrudes from the vacuum vessel and is used for the room temperature joint. The last 60 mm of the lead extends into a bath of superfluid helium at 2.0 K, where it is bonded to superconducting wire.

There are six total current leads for three magnets that carry an electrical current up to 50 A when the magnets are energized. XFEL cryomodules have two thermal shield circuits, with the outer circuit operating between 40 K and 80 K and the inner circuit operating between 5 K and 8 K. The current leads are thermally intercepted by using copper clamps brazed to copper braiding. The current leads fit in grooves in the clamps to provide maximum surface area contact between the clamp and lead. The outer intercept clamps the leads between 0.42 m and 0.49 m from the start of the brass core. The inner intercept clamps the lead between 0.46 and 0.53 m from the end of the brass core.

**ANALYTICAL MODEL**

Since the length of the current leads is orders of magnitude larger than the diameter, a simple approach to estimate heat loads is to analyze the leads as a one-dimensional thermal conduction problem with internal heat generation. The heat transfer rate at any point along a one-dimensional conductor can be determined using Eq. 1 [3]:

$$\dot{Q} = I^2 R_{e,eff} \left( \frac{x}{L} - \frac{1}{2} \right) + \frac{T_h - T_c}{R_{t,eff}} \tag{1}$$

where $\dot{Q}$ is the heat conducted through the lead at distance $x$ from $T_h$ towards $T_c$, $I$ is the electrical current, $R_{e,eff}$ is the effective electrical resistance, $L$ is the total conduction length, $T_h$ is the hot end temperature, $T_c$ is the cold end temperature, and $R_{t,eff}$ is the effective thermal resistance. Current leads are often made with one or more layers of material bonded to or otherwise surrounding a core material. Each of the layers contributes to the effective thermal resistance, which can be estimated using the formula in Eq. 2 for parallel thermal resistances with identical conduction lengths:

$$R_{t,eff} = \frac{L}{k_{av,1}A_1 + k_{av,2}A_2 + .... + k_{av,n}A_n} \quad (2)$$

where $k_{av}$ is the integrated average thermal conductivity of the material over the temperature range $T_h$ to $T_c$, $A$ is the material cross sectional area, and the subscripts *1* through *n* refer to each of the different materials. The current leads may also have multiple electrical conductors. The effective electrical resistance can be estimated using the formula in Eq. 3 for parallel electrical resistances with identical conduction lengths:

$$R_{e,eff} = \frac{L}{A_1/\rho_{av,1} + A_2/\rho_{av,2} + .... + A_n/\rho_{av,n}} \quad (3)$$

where $\rho_{av}$ is the integrated average thermal conductivity of the material over the temperature range $T_h$ to $T_c$. Eqs. 1-3 fail to account for two major aspects of the PXFEL2 current leads, with the first being the varying copper cross sectional area and the second being the use of thermal intercepts. These factors can be addressed by breaking the leads up into three sections along the axial length at the midpoint of the intercepts. Each section can then be analyzed independently and energy balances can be used to determine the net heat loads.

Shown in Table 1 are the heat loads predicted by Eqs. 1-3 for a single current lead. Section A represents the lead from the vacuum vessel to the outer shield intercept, Section B represents the lead between the intercepts, and Section C is from the inner shield intercept to the helium bath. The intercept temperatures in Table 1 are nominally the same as the experimentally measured temperatures on the lead clamps for PXFEL2. The small length of electrical conducting core extending from either side of the stainless steel tube for soldering was neglected.

All material properties were tabulated from CRYOCOMP [4] and all helium properties are from HEPAK [5]. All equations were solved with temperature averaged properties using the iterative equation solver EES [6]. Copper conductivity is highly dependent on purity at cryogenic temperatures and the purity is typically defined by the Residual Resistance Ratio (RRR). The copper purity was examined for two conditions: a RRR of 120 and a RRR of 300.

**TABLE 1. Predicted heat loads for a single current lead using analytical model (RRR=120 and RRR=300)**

| | Inputs for Eq. 1-3 | | | Resistances | | $I = 0$ [A] | | $I = 50$ [A] | | Cold End Net Heat Load | | |
|---|---|---|---|---|---|---|---|---|---|---|---|---|
| Section | $T_h$ [K] | $T_c$ [K] | $L$ [m] | $R_{t,eff}$ [K/W] | $R_{e,eff}$ [µΩ] | $\dot{Q}(x=0)$ [W] | $\dot{Q}(x=L)$ [W] | $\dot{Q}(x=0)$ [W] | $\dot{Q}(x=L)$ [W] | Static [W] | Total [W] | Dynamic [W] |
| *RRR=120* | | | | | | | | | | | | |
| A | 300 | 60 | 0.435 | 112 | 473 | 2.15 | 2.15 | 1.56 | 2.74 | 1.74 | 2.44 | 0.703 |
| B | 60 | 10 | 0.330 | 122 | 89.8 | 0.412 | 0.412 | 0.299 | 0.524 | 0.271 | 0.393 | 0.122 |
| C | 10 | 2 | 0.435 | 56.7 | 8.23 | 0.141 | 0.141 | 0.131 | 0.151 | 0.141 | 0.151 | 0.010 |
| *RRR=300* | | | | | | | | | | | | |
| A | 300 | 60 | 0.435 | 110 | 469 | 2.18 | 2.18 | 1.59 | 2.77 | 1.58 | 2.25 | 0.68 |
| B | 60 | 10 | 0.330 | 82.6 | 69.2 | 0.605 | 0.605 | 0.518 | 0.692 | 0.261 | 0.352 | 0.091 |
| C | 10 | 2 | 0.435 | 23.2 | 3.30 | 0.344 | 0.344 | 0.340 | 0.349 | 0.344 | 0.349 | 0.005 |

## NUMERICAL MODEL

The analytical model is simple and easy to use, but has several limitations. The analytical model does not consider the thermal resistance across the helium gas filled gap between the Kapton insulator and the stainless steel tube. The thermal resistance of the intercepts between the flowing fluid and the current lead are completely neglected by the analytical model. Pure materials such as copper can have large thermal and/or electrical property changes from small temperature changes; therefore, using integrated average properties across a section of the current lead may lead to inaccuracies.

Numerical models are not as simple to create as analytical models but do have the advantage of greatly increased flexibility for modeling heat transfer with complicated geometry and temperature dependent material and fluid properties. The numerical model for the current leads is analogous to a heat exchanger model [7] in parallel flow, since both inner conductor and outer sheath conduct heat in the same direction. The current leads can be broken up into an arbitrarily large number of nodes. The electrical and thermal resistance between each node can be calculated, using Eq. 2 and Eq. 3 respectively, by assuming an $L$ equal to the distance between each node.

The gap between the electrical conductor and the stainless steel tubing sheath acts much like a heat exchanger wall. Any heat that is removed by the thermal intercepts from the highly conductive inner conductor must cross the low pressure helium filled gap. Axial conduction through the helium is neglected, since the conductivity of helium is far less than of copper, brass, and stainless steel. The cold end of the lead is exposed to a He II bath, but calculations show that the distance that the He II could creep through the gap is negligible. Thermoacoustic oscillations are notoriously hard to predict and were not addressed by the numerical model.

The heat generated by ohmic dissipation in the electrical conductor is treated similar to a parasitic heat load in the heat exchanger. A pictorial representation of the 2-D finite difference model used for the current leads is shown in Fig. 1. The subscript "ic" is for inner conductor, "oc" is for outer conductor, "gen" is for heat generation, and "int" is for the heat removed by the intercepts.

The heat removal of the thermal intercepts can be included in the lead model similar to how a parasitic heat load is included in a heat exchanger model. The only difference is a sign change since heat flows in the reverse direction. For all nodes that do not touch an intercept clamp $Q_{int}$ is set equal to zero. The resistance network used to model the nodes is shown in Fig.

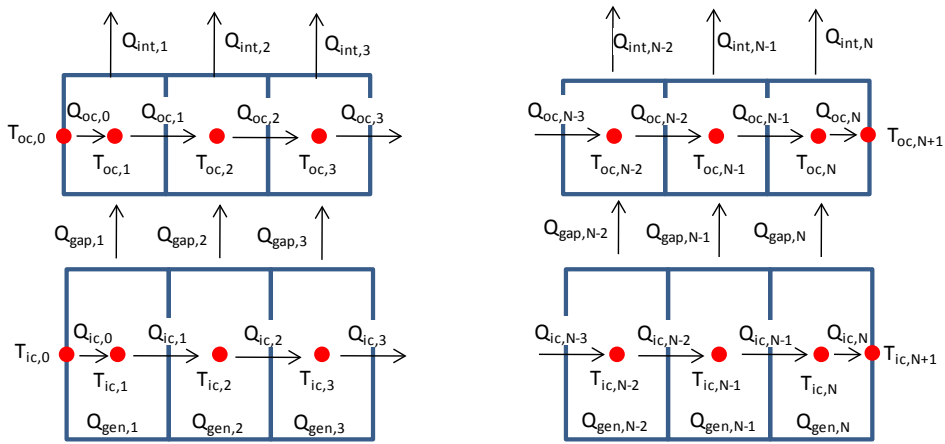

**FIGURE 1.** Finite difference network used to model the current leads

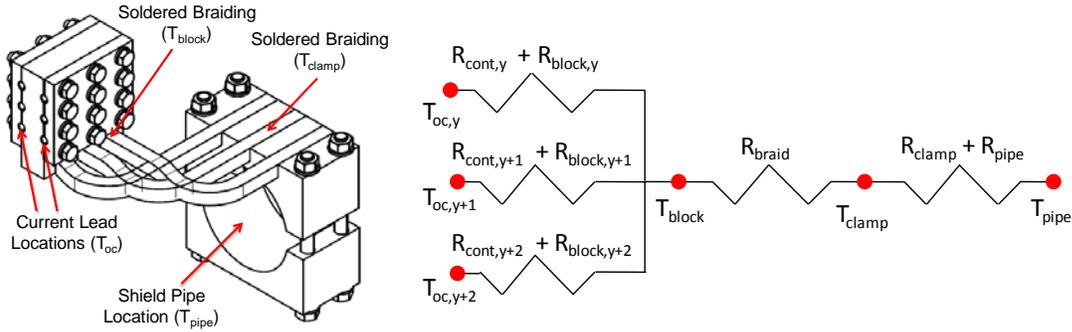

**FIGURE 2.** Drawing and resistance network for the thermal intercepts

2, with node $y$ being the first node within the current lead clamp. There is a contact resistance $R_{cont}$ between the lead outer sheath and the clamp surface. The current lead clamp is machined from a copper block and will also have a small resistance $R_{block}$. The bottom surface of the block where the copper braiding is soldered is at temperature $T_{block}$ and the braids have a resistance $R_{braid}$. The braids terminate at a solder joint on a copper pipe clamp that is at a temperature $T_{clamp}$. There is a resistance across the pipe clamp $R_{clamp}$ and a resistance across the pipe wall to the bulk fluid $R_{pipe}$. The shield coolant has an average temperature of $T_{pipe}$.

The thermal conductivity of copper is highly sensitive to its purity and is one of the leading sources of uncertainty regarding the overall intercept thermal and electrical resistance. The exact purity of the copper used in the blocks and braiding is unknown, so the RRR was estimated to be 50. Experiments have shown that there is a fair amount of thermal resistance between the thermal shield piping and the lead clamps, so the actual RRR is expected to be fairly low.

The braids are soldered into grooves on the clamps, so the braided joints are not expected to have a large thermal resistance. The pipe clamp has a layer of indium between the clamp and the pipe; therefore, the pipe clamp joint is not expected to have a large thermal resistance. However, there is no intermediate material between the stainless steel outer tubing on the current leads and the copper clamp block. Thermal contact resistances are primarily a function of the contact force and not the contact pressure, which means that the contact surface area is not the dominant factor [8]. The current leads sit in grooves that are slightly too small for the stainless steel tubing, which ensures that a large force is required to bring the copper blocks together due to the distortion of the stainless steel tubing. Salermo and Kittel's literature review of cryogenic joints showed that joints typically have a conductance around 0.02 K/W [8], but with higher applied forces the conductance could be as high as 1.5 K/W. The applied forces were higher for the current leads than any of the values cited in the literature review, so a value of 1.5 K/W was chosen for the model.

The RRR of the copper used as an electrical conductor is expected to be about 120. The outer shield inlet temperature is generally about 40 K and the inner shield inlet temperature is generally about 4.5 K. The temperature profile along the length of the lead is shown in Fig.3 for both the minimum current ($I=0$ A) and maximum current ($I=50$ A). The locations near the intercepts are shown in detail in Fig. 4. One important point to note is that the changes in copper thickness have a much larger effect on the temperature gradient than the thermal intercepts.

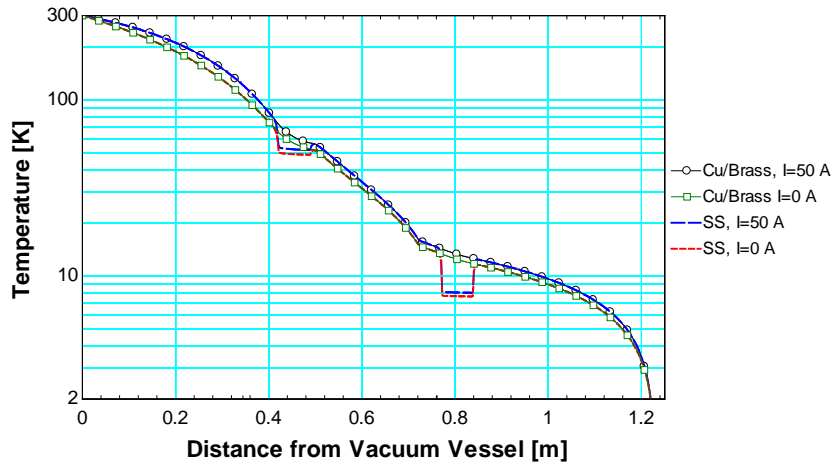

**FIGURE 3. Calculated temperature profile a RRR=120 current lead with a temperatures of 40 K for the outer shield fluid and 4.5 K for the inner shield fluid at I=0 A and I=50 A.**

A parametric study was performed using the numerical model. Three variables were varied: the purity of the copper (RRR=120 or RRR=300), current (0 A min, 50 A max), and outer shield temperature (40 K or 80 K). The results are shown in Table 3. The resistances of the braids were typically in the range of 5 to 10 K/W. A couple conclusions can be drawn from Table 2. The first is that using $LN_2$ (T=80K) rather than 40 K helium will roughly double the dynamic 2 K bath heat load and will increase the static heat load by almost 50%. Using high purity copper has the tendency to increase the static heat load far more than the dynamic heat load is reduced. The second is that the increase in static heat load dominates the reduction in dynamic heat load when higher purity copper is used. The only case where high purity copper should be considered for use is in a current lead with high currents and high duty factors.

## CMTB HEAT LOADS

The static heat loads of the current leads are difficult to determine since even a well-designed test cryostat will still have a static heat load on the helium bath regardless of whether current leads are present. However, the dynamic heat loads can be readily determined by measuring the increase in flow rates due to the increase in the electrical power dissipated in the leads. The experimentally measured dynamic heat loads measured on CMTB for PXFEL2 are shown in Table 3. The cryomodule SRF cavities were unpowered for these measurements. The fluid temperatures were kept almost constant and the mass flow rate was varied.

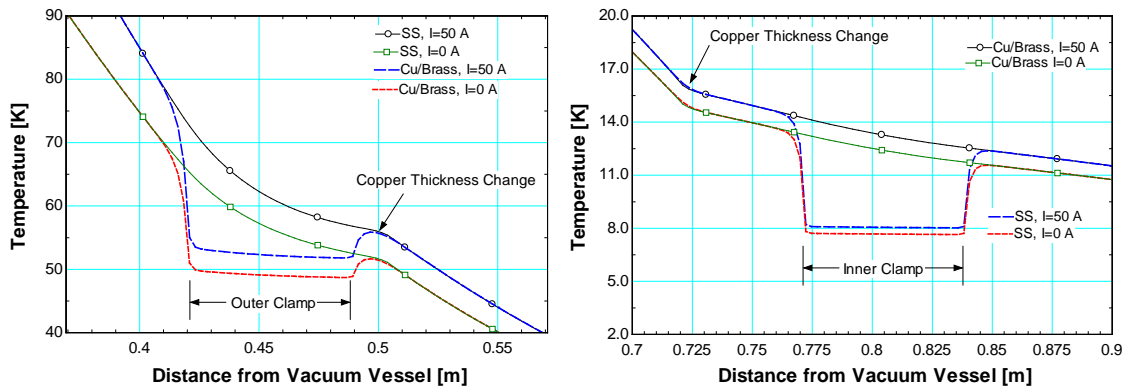

**FIGURE 4. Closeup view of the intercept locations in Fig. 3. Segments of the current lead near the outer intercept and inner intercept are shown on the left and right, respectively.**

**TABLE 2.** Numerical model parametric study results for a single current lead. The inner shield fluid temperature was assumed constant at 4.5 K.

| Section | RRR [-] | $T_{pipe,out}$ [K] | $I$ [A] | $Q_{int,out}$ [W] | $Q_{int,in}$ [W] | $Q_{bath}$ [W] |
|---|---|---|---|---|---|---|
| Static  | 120 | 40 | 0  | 1.66 | 0.27 | 0.22 |
| Total   | 120 | 40 | 50 | 2.16 | 0.31 | 0.27 |
| Dynamic |     |    |    | 0.50 | 0.04 | 0.05 |
| Static  | 120 | 80 | 0  | 1.08 | 0.38 | 0.30 |
| Total   | 120 | 80 | 50 | 1.90 | 0.47 | 0.38 |
| Dynamic |     |    |    | 0.72 | 0.09 | 0.08 |
| Static  | 300 | 40 | 0  | 1.49 | 0.24 | 0.49 |
| Total   | 300 | 40 | 50 | 1.92 | 0.25 | 0.52 |
| Dynamic |     |    |    | 0.43 | 0.01 | 0.03 |
| Static  | 300 | 80 | 0  | 0.89 | 0.30 | 0.60 |
| Total   | 300 | 80 | 50 | 1.61 | 0.34 | 0.66 |
| Dynamic |     |    |    | 0.72 | 0.04 | 0.06 |

The pipe location where the intercept is clamped is much closer to the cryomodule inlet temperature than the outlet temperature. The shield flow rates were measured using cold orifice flowmeters and the bath flow rate was measured using a thermal flowmeter at the discharge of the vacuum pump. The temperatures of the clamps on the process piping and the current leads were measured by clamped surface mounted sensors using uncured epoxy to improve the thermal contact between the sensor and surface. More information on CMTB heat load measurement methods can be found in the literature [9-11].

## XMTS HEAT LOADS

The current leads testing results in XMTS are shown in Table 4. The lead diameters, lengths and clamps are identical to those used in the PXFEL2 cryomodule; however, the bending locations are a little different due to space constraints within the XMTS cryostat. The flow rate was measured using a cryogenic coriolis flowmeter, which was previously verified by room temperature flowmeters at the discharge of the vacuum pump.

**TABLE 3:** Measured temperatures and dynamic heat loads for all six current leads of the PXFEL2 cryomodule in CMTB at DESY

| Location | Current [A] | Inlet [K] | Outlet [K] | Pipe Clamp [K] | Lead Clamp [K] | Dynamic Heat Load [W] |
|---|---|---|---|---|---|---|
| Outer Shield | 0.0 | 40.2 | 46.5 | 54.1 | 58.4 | 2 to 5 |
| Outer Shield | 50.0 | 41.2 | 47.6 | 59.9 | 68.5 | |
| Inner Shield | 0.0 | 4.3 | 4.8 | 9.3 | 10.0 | 0.9 |
| Inner Shield | 50.0 | 4.4 | 4.8 | 9.8 | 10.7 | |
| Helium Bath[1] | 0.0 | 2.0 | 2.0 | - | - | 0.0 |
| Helium Bath[1] | 50.0 | 2.0 | 2.0 | - | - | |
| [1]The leads terminate in a 2K bath so outlet and clamp temperatures not applicable |||||||

**TABLE 4:** Measured 2.0 K helium bath dynamic heat load for all six PXFEL2 current leads in XMTS.

| Current [A] | Helium Mass Flow [g/s] | Dynamic Heat Load [W] |
|---|---|---|
| 0 | 0.237 | 0.000 |
| 10 | 0.241 | 0.093 |
| 20 | 0.251 | 0.326 |
| 30 | 0.272 | 0.816 |
| 40 | 0.305 | 1.584 |
| 50 | 0.340 | 2.400 |

## CONCLUSION

Unfortunately the data from XMTS and CMTB are not in good agreement. The experimental test facilities were designed to measure large heat loads. The lead dynamic loads are small and have nearly the same magnitude as the noise in the measurements. The relative uncertainty of the measured loads is quite high. Another possibility for the discrepancy is due to geometry. The same length leads are used in both test facilities, but the leads must be bent differently to fit into XMTS. DESY measurements for other current leads that are similar to the PXFEL2 leads have also yielded inconsistent results. One possibility is the purity of the copper changes greatly from sample to sample due to inconsistent electroplating. Another possibility is that the design is susceptible to thermoacoustic oscillations from small changes in geometry. The numerical model can neither be validated nor disproved until further data is taken.

## ACKNOWLEDGEMENTS

Special thanks to Arkadiy Klebaner of FNAL and Bernd Petersen of DESY for organizing this collaborative project. Work supported by Fermi Research Alliance, LLC under Contract No. DE-AC02-07CH11359 with the United States Department of Energy and by DESY.